Reduced hierarchy equations of motion approach with Drude plus Brownian spectral distribution: Probing electron transfer processes by means of two-dimensional correlation spectroscopy

Yoshitaka Tanimura

*Department of Physics, University of Hamburg, Centre for Free Electron Laser Science, DESY, Notkestrasse 85, D-22607 Hamburg, Germany*
*Department of Chemistry, Graduate School of Science, Kyoto University, Kyoto 606-8502, Japan*

We theoretically investigate an electron transfer (ET) process in a dissipative environment by means of two-dimensional (2D) correlation spectroscopy. We extend the reduced hierarchy equations of motion approach to include both overdamped Drude and underdamped Brownian modes. While the overdamped mode describes the inhomogeneity of a system in the slow modulation limit, the underdamped mode expresses the primary vibrational mode coupled with the electronic states. We outline a procedure for calculating 2D correlation spectrum that incorporates the ET processes. The present approach has the capability of dealing with system-bath coherence under an external perturbation, which is important to calculate nonlinear response functions for non-Markovian noise. The calculated 2D spectrum exhibits the effects of the ET processes through the presence of ET transition peaks along the $\Omega_1$ axis, as well as the decay of echo signals.

I. INTRODUCTION

Quantum coherence and its destruction by coupling to a dissipative environment plays an important role in time resolved optical response[1,2] as well as nonadiabatic electron transfer (ET)[3,4,5,6,7] in condensed phases. Each of these processes involves coupling between the internal vibrations or electronic excitations of a molecule and the external degrees of freedom of its environment.[8,9] Femtosecond spectroscopy provides a direct means for studying nuclear dynamics in the condensed phase.[10,11,12] Since the spectral lines for these processes are often broadened and appear in similar positions, it is not easy to explore their roles with linear spectroscopy. This difficulty can be overcome by ultrafast nonlinear spectroscopies involving many laser interactions such as pump–probe spectroscopy.[13,14,15] These techniques make it possible to utilize more than one time-evolution period and allow us to distinguish dynamical processes with different time responses. Recently, two-dimensional electronic spectroscopy (2DES) has also taken part in the investigation of the dynamics of exciton transfer[16,17,18,19] and electron transfer processes,[20,21] which stimulated the investigation of this field especially focusing on a role of quantum coherence.[22,23,24] In 2D spectroscopies, the multi-body correlation function of a transition dipole as a function of the time durations between the pulses is measured using ultra short pulses. A two-dimensional contour map of the signals in a Fourier space can unveil the exciton-exciton interactions, dephasing, and relaxation processes that are usually hidden by the broadening of spectrum in linear spectroscopy.[25]

In a widely used model for time resolved optical responses as well as ET, the electronic states are coupled to an intermediate harmonic nuclear or intramolecular vibrational mode, which is in turn coupled to a heat-bath.[8,9] 2D spectroscopies obtained from such systems may provide useful information especially for relaxation and dephasing processes. However, its theoretical analysis is much more complex compared with lower order processes. If ET coupling does not exist, the optical response function approach based on a perturbative expansion of the optical polarization in powers of the laser fields can be used to study 2D spectroscopies even for a general spectral distribution with a strong system-bath coupling case.[2,26,27,28] We may handle the ET processes in a similar manner as the optical transition if ET coupling is weak,[8,9,29,30] but an extension of the response function approach to the case of strong ET coupling is not easy.

Alternatively, optical processes can be calculated using a direct integration of the equations of motion in the presence of ET coupling and external fields. In the absence of dissipation, quantum ET transitions can be studied by a wide variety of numerical methods based on the relevant wave function.[31] When dissipation is important, a reduced density matrix has to be used in the presence of the bath in order to study the irreversibility of system dynamics toward the thermal equilibrium state.[32] A difficulty with this approach is in the treatment of the dissipation processes induced by a heat bath. These are usually incorporated by using equations of motion for a reduced density matrix such as the Redfield equation,[33,34,35,36] and the stochastic Liouville equation.[37,38]

The Redfield (or master) equation approach requires several assumptions, such as the rotating wave approximation, the white-noise (van Hove) approximation, and a factorized initial condition. Beyond the limitations of these approximations, the equations of motion of this type sometime produce unphysical results such as a negative probability of density matrix elements. For the master equation, this phenomenon is known as breaking of dynamic positivity. This is the limitation of some of the reduced equation of motion approaches. If one modifies the interaction in the resonant form or the rotating wave approximation form which leads to the Lindblad form of the master equation, the positivity problem does not become apparent, and the dynamics of the system might be different compared to the real system.[39] To have physically meaningful results, one has to maintain the conditions to satisfy the approximations. Although the time-convolution-less (TCL) form of the Redfield equation can handle non-Markovian noise fairly well for the case where the system Hamiltonian and the system-bath interaction Hamiltonians are commute,[40,41] its applicability is still limited because it cannot handle the system-bath coherence over the external laser interaction, which plays a major role in 2D spectroscopies.[42]

While the stochastic Liouville equation can handle non-Markovian noise, its applicability is strongly limited.[32] This is because the stochastic theory is phenomenological and does not ensure the thermal equilibrium state at finite temperature and it also has to utilize a resolvent in a continued fractional form to calculate physical observables.[37,38]

To eliminate all of the above mentioned limitations, one can derive the hierarchy equations of motion (HEOM) for the reduced density matrix derived from a system-bath Hamiltonian.[32,43] Since HEOM approach is a dynamical theory based on the Hamiltonian, the system approaches a thermal equilibrium state at finite temperature, when the external perturbation is switched off. Thanks to the truncation schemes for higher-order hierarchy elements[44,45,46,47] one can numerically integrate HEOM for variety of systems expressed as Wigner distributions[48,49,50,51,52] and energy eigen states[53,54,55,56,57] as well as when a time-dependent external perturbation such as laser[49] or magnetic excitation[58] is present. By generalizing hierarchy structures, one can deal with a low temperature system[45,59,60] as well as general spectral distributions[61] including Brownian spectral distributions[62,63,64,65] and a Lorentzian distribution.[66,67] This formalism is valuable since it can handle not only strong system-bath coupling, but also quantum coherence between the system and bath, which plays important roles in multidimensional spectroscopy,[50-55,67,68,69] energy transfer processes in photosynthetic antenna systems[70,71,72,73,74,75,76,77] and DNA systems,[78] ET process,[49,79] and processes discussed in a quantum information theory.[80,81,82,83]

While most research with the HEOM approach assumed the Drude spectral distribution, we have shown that the ET problem can be handled in a nonperturbative manner for both the system-bath and ET couplings by applying the hierarchy formalism to the Brownian oscillator (BO) spectral distribution that arises from the canonical transformation of ET system.[62,63,64] Because realistic environments in many cases involve both the overdamped Drude and underdamped Brownian modes as shown by molecular dynamics simulations,[84,85,86,87,88] an extension to the multimode case is necessary. In this paper, we demonstrate a way to deal with the Drude+BO spectral distribution in the framework of HEOM formalism. Moreover, we calculate 2D correlation spectrum for a case that both ET and optical transitions become important to investigate the role of dissipation in coherent spectroscopies involving ET processes.

The organization of the paper is as follows: In Sec. II we present a model Hamiltonian for ET transition problem. In Sec. III, we derive reduced hierarchy equations of motion for the Drude plus Brownian oscillator mode. In Sec. IV, we explain a procedure for calculating two-dimensional correlation spectra. In Sec. V, the numerical results of linear absorption spectra and 2D spectra for different ET coupling parameters are shown and discussed. Section VI is devoted to concluding remarks.

## II. MODEL

In a widely used model for time resolved optical responses as well as ET, the electronic states are coupled to intermediate harmonic nuclear or intramolecular vibrational modes, which are in turn coupled to a heat-bath.[8,9] The primary electronic system *A* is taken to be a two-level system with a lower sate $|0\rangle$ and an upper state $|1\rangle$. The two states interact through an ET coupling parameter $\Delta$ and a laser interaction $f(t)$. The two-level system is in turn coupled to harmonic vibrational modes. The molecular Hamiltonian is then expressed as[5,6,7,8,9,29,30]

$$\hat{H}_{A+O} = |0\rangle H_0(\mathbf{p},\mathbf{q})\langle 0| + |1\rangle H_1(\mathbf{p},\mathbf{q})\langle 1|$$
$$+ \left(\frac{1}{2}\hbar\Delta + f(t)\right)(|1\rangle\langle 0| + |0\rangle\langle 1|), \quad (2.1)$$

where

$$H_0(\mathbf{p},\mathbf{q}) = \sum_j \left[\frac{p_j^2}{2m_j} + \frac{m_j \omega_j^2}{2}\left(q_j - \frac{1}{2}d_j\right)^2\right] - \frac{1}{2}\hbar\omega_0, \quad (2.2)$$

$$H_1(\mathbf{p},\mathbf{q}) = \sum_j \left[\frac{p_j^2}{2m_j} + \frac{m_j \omega_j^2}{2}\left(q_j + \frac{1}{2}d_j\right)^2\right] + \frac{1}{2}\hbar\omega_0, \quad (2.3)$$

and $p_j$, $q_j$, $m_j$, $\omega_j$, and $d_j$ represent, respectively, the momentum, coordinate, mass, frequency, and displacement of the *j*th nuclear degrees of freedom strongly coupled to the electronic state. A schematic view of the system for a single mode case is depicted in Fig. 1. The nuclear oscillator modes further coupled to the harmonic bath systems are expressed as

$$H_B = \sum_j \sum_{n_j} \left[\frac{p_{n_j}^2}{2m_{n_j}} + \frac{m_{n_j}\omega_{n_j}^2}{2}\left(x_{n_j} - q_j\right)^2\right], \quad (2.4)$$

where $x_{n_j}$, etc are the bath oscillator variables for the *j*th nuclear mode. We can reduce the nuclear oscillator mode by performing a canonical transformation of the oscillator + bath coordinates. After the transformation, the total Hamiltonian is reduced to[8]

$$\hat{H}_{tot} = \hat{H}_A + \hat{V}\sum_j \sum_{n'_j} c_{n'_j} x_{n'_j} + \sum_j \sum_{n'_j}\left[\frac{p_{n'_j}^2}{2m_{n'_j}} + \frac{m_{n'_j}\omega_{n'_j}^2}{2}x_{n'_j}^2\right], \quad (2.5)$$

where $\hat{V} = \hat{\sigma}_z/2$,

$$\hat{H}_A = \frac{1}{2}\hbar\omega_0\hat{\sigma}_z + \left(\frac{1}{2}\hbar\Delta + f(t)\right)\hat{\sigma}_x, \tag{2.6}$$

and $\hat{\sigma}_i$ ($i = x, y, z$) are the Pauli matrices. If we assume the spectral density of the oscillator-bath to be $J_j(\omega) = \gamma_j\omega$, then the spectral density of Eq.(2.5) becomes

$$J'_j(\omega) = \frac{2\hbar\lambda_j}{\pi}\frac{\gamma_j\omega_j^2\omega}{\left(\omega_j^2 - \omega^2\right)^2 + \gamma_j^2\omega^2}, \tag{2.7}$$

where

$$\lambda_j = \frac{m_j d_j^2 \omega_j^2}{2\hbar}. \tag{2.8}$$

Note that the above spectral distribution effectively reduces to the Drude form for $\gamma_j \gg \omega_j$ as [6,2]

$$J_D(\omega) = \frac{2\hbar\lambda_D}{\pi}\frac{\gamma_D\omega}{\gamma_D^2 + \omega^2}. \tag{2.9}$$

We consider one overdamped mode and one underdamped mode to model electron transfer process in a solvated or protein environment. While the underdamped mode represents a vibrational mode, the overdamped mode represents an inhomogeneity of the system in the slow modulation limit.[26,27,28,32] Such example involves a large dye molecule with two electronic states (the ground state and an excited sate or two excited states).[7] Note that an optical metal-metal charge transfer (MMCT) system in a solvated environment[13,14,15] may be described in a similar framework, although MMCT is described by the free energy potential surface while the present model is described by the potential energy surface.

## III. REDUCED HIERACHY EQUATIONS OF MOTION FOR DRUDE+BROWNIAN BATH

After the bath degrees of freedom are traced out, the reduced density matrix element is expressed in the path integral form as[3,2]

$$\rho(\bar{\psi},\psi';t) = \int D[Q(\tau)] \int D[Q'(\tau)] e^{\frac{i}{\hbar}S_A[Q;t]} \left( \prod_j F_j[Q,Q';t] \right) e^{-\frac{i}{\hbar}S_A[Q';t]}, \quad (3.1)$$

where $Q(\tau) = \{\psi(\tau), \bar{\psi}(\tau)\}$ and $Q'(\tau) = \{\psi'(\tau), \bar{\psi}'(\tau)\}$ are the coherent state representation of sets of Grassmann numbers that describe the states of the system, $|0\rangle$ and $|1\rangle$ and $\int D[Q(\tau)]$ represents the functional integral. The action for the system's Hamiltonian $H_A$ is denoted by $S_A[Q;t]$. The bath effects are described by the Feynman-Vernon influence functional. For the distribution Eq. (2.7), the influence functional is calculated as

$$F_j[Q,Q';t] = e^{-\frac{i}{\hbar}\int_{t_0}^{t} ds \int_{t_0}^{s} du \sigma_z^\times(s)\left(-iL_1^j(s-u)\sigma_z^\circ(u) + L_2^j(s-u)\sigma_z^\times(u)\right)}, \quad (3.2)$$

where $\sigma_z^\times(s) \equiv \sigma_z(Q(s)) - \sigma_z(Q'(s))$ and $\sigma_z^\circ(s) \equiv \sigma_z(Q(s)) + \sigma_z(Q'(s))$ are the commutator and anticommutator expressed in the Grassmann variables, respectively. The time-dependent kernels corresponding to the fluctuation $iL_1^j(t) = \langle [q_j(t), q_j] \rangle / \hbar$ and the dissipation $L_2^j(t) = \langle \{q_j(t), q_j\} \rangle / 2\hbar$ are expressed by the spectral distribution as $iL_1^j(t) = \int_0^\infty d\omega J_j'(\omega) \sin(\omega t)$ and $L_2^j(t) = \int_0^\infty d\omega J_j'(\omega) \cos(\omega t) \coth(\beta \hbar \omega / 2)$, respectively.

In this paper, we deal with two modes, one with an overdamped oscillator and the other with an underdamped oscillator. The ET coupling and the energy difference between the two potentials are chosen to be the same and denoted by $\Delta$ and $\omega_0$, respectively. As illustrated in Fig. 2, we may consider two cases. Namely, (a) the two oscillators are independently coupled to their own bath or (b) one oscillator coupled to two baths. If the frequencies of the two oscillators in the case (a) are the same, the cases (a) and (b) become identical because the system-oscillator-bath interactions are linear. Since the frequency of the oscillator does not play a role in the overdamped mode, the case (a) and (b) become identical. We thus consider the spectral distribution expressed as

$$J'(\omega) = \frac{2\hbar\lambda_o}{\pi}\frac{\gamma_o\omega}{\gamma_o^2+\omega^2} + \frac{2\hbar\lambda_u}{\pi}\frac{\gamma_u\omega_u^2\omega}{\left(\omega_u^2-\omega^2\right)^2+\gamma_u^2\omega^2}. \quad (3.3)$$

The correlation functions are then calculated as[61-64]

$$iL_1(t) = \frac{i\lambda_o\gamma_o}{2}e^{-\gamma_o t} - \frac{i\lambda_u\omega_u^2}{2\zeta_u}\left(e^{-\left(\frac{\gamma_u}{2}-i\zeta_u\right)t} - e^{-\left(\frac{\gamma_u}{2}+i\zeta_u\right)t}\right), \quad (3.4)$$

$$L_2(t) = \frac{\lambda_o}{2\omega_0}\cot\left(\frac{\beta\hbar\gamma_o}{2}\right)e^{-\gamma_o t} - \frac{\lambda_u}{2\zeta_u}\left[A_u^- e^{-\left(\frac{\gamma_u}{2}-i\zeta_u\right)t} - A_u^+ e^{-\left(\frac{\gamma_u}{2}+i\zeta_u\right)t}\right]$$
$$-\sum_{k=1}^{\infty}\left(\frac{2\lambda_o\gamma_o}{\beta\hbar}\frac{\nu_k}{\nu_k^2-\gamma_o^2} + \frac{4\lambda_u\gamma_u\omega_u^2}{\beta\hbar}\frac{\nu_k}{\left(\omega_u^2+\nu_k^2\right)^2-\gamma_u^2\nu_k^2}\right)e^{-\nu_k t}, \quad (3.5)$$

where $\zeta_u = \sqrt{\omega_u^2-\gamma_u^2/2}$ and $A_u^\pm = \coth\left(\beta\hbar i\left(\gamma_u\pm 2i\zeta_u\right)/4\right)$. The reduced hierarchy equations of motion (HEOM) can be obtained by considering the time derivative of the reduced density matrix with the kernel eqs.(3.4) and (3.5). The procedures are parallel to Refs. 32, 45, 63, 64. We denote the number of the hierarchy elements for $\gamma_u\pm 2i\zeta_u$ as $m_\pm$ and the number of $k$th Matsubara frequencies by $j_k$. The cutoff of Matsubara frequencies is expressed by $K$. The hierarchy equations of motion for Drude+ Brownian spectral distribution is then expressed as

$$\dot{\hat{\rho}}_{j_1\cdots j_K}^{(n,m_-,m_+)}(t) = -\left[\frac{i}{\hbar}\hat{H}_A^\times - \gamma^{(n,m_-,m_+)} + \sum_{k=1}^K j_k \nu_k - \hat{\Xi}^{DB}\right]\hat{\rho}_{j_1\cdots j_K}^{(n,m_-,m_+)}(t)$$
$$+\hat{V}^\times\left[\hat{\rho}_{j_1\cdots j_K}^{(n+1,m_-,m_+)}(t) + \hat{\rho}_{j_1\cdots j_K}^{(n,m_-+1,m_+)}(t) + \hat{\rho}_{j_1\cdots j_K}^{(n,m_-,m_++1)}(t)\right]$$
$$+n\hat{\Theta}\hat{\rho}_{j_1\cdots j_K}^{(n-1,m_-,m_+)}(t) + m_-\hat{\Theta}_-\hat{\rho}_{j_1\cdots j_K}^{(n,m_--1,m_+)}(t) + m_+\hat{\Theta}_+\hat{\rho}_{j_1\cdots j_K}^{(n,m_-,m_+-1)}(t)$$
$$+\sum_{k=1}^K \hat{V}^\times \hat{\rho}_{j_1,\cdots,j_k+1,\cdots j_K}^{(n,m_-,m_+)}(t) + \sum_{k=1}^K j_k\nu_k\hat{\Psi}_k^{DB}\hat{\rho}_{j_1,\cdots,j_k-1,\cdots j_K}^{(n,m_-,m_+)}(t), \qquad (3.6)$$

where

$$\gamma^{(n,m_-,m_+)} = n\gamma_o + \frac{(m_- + m_+)\gamma_u}{2} - i(m_- - m_+)\zeta_u, \qquad (3.7)$$

$$\hat{\Theta} = \frac{\lambda_o}{2\omega_0}\left[\hat{V}^\circ + i\cot\left(\frac{\beta\hbar\gamma_o}{2}\right)\hat{V}^\times\right], \qquad (3.8)$$

$$\hat{\Theta}_\pm = \frac{\lambda_u\omega_u^2}{2\zeta_u}\left\{\mp\hat{V}^\circ \pm A_u^\mp\hat{V}^\times\right\}, \qquad (3.9)$$

$$\hat{\Psi}_k^{DB} = \left(\frac{2\lambda_o\gamma_o}{\beta\hbar}\frac{\nu_k}{\nu_k^2 - \gamma_o^2} + \frac{4\lambda_u\gamma_u\omega_u^2}{\beta\hbar}\frac{\nu_k}{\left(\omega_u^2 + \nu_k^2\right)^2 - \gamma_u^2\nu_k^2}\right)\hat{V}^\times, \qquad (3.10)$$

and $\hat{\Xi}^{DB} = \hat{V}^\times\sum_{k=K+1}^\infty \hat{\Psi}_k^{DB}$. For the condition $\gamma^{(n,m_-,m_+)} + \sum_{k=1}^K j_k\nu_k \gg \max\{\omega_0, \Delta\}$, this infinite hierarchy can be truncated by the terminator as

$$\dot{\hat{\rho}}_{j_1\cdots j_K}^{(n,m_-,m_+)}(t) \approx -\left[\frac{i}{\hbar}\hat{H}_A^\times - \gamma^{(n,m_-,m_+)} + \sum_{k=1}^K j_k\nu_k - \hat{\Xi}^{DB}\right]\hat{\rho}_{j_1\cdots j_K}^{(n,m_-,m_+)}(t). \qquad (3.11)$$

Since both the Drude and Brownian modes can share the Matsubara frequency expansion terms, the increase in the number of hierarchies is moderate. The total number of hierarchy elements is evaluated as $L_{tot} = (N + M + K +1)/(K +1)!/(N + M)!$, while the total number of termination elements is $L_{term} = (N + M + K)/K!/(N + M)!$, where $N$ is the depth of hierarchy for $n$, $m_+$ and $m_-$, and $M$ is the increase of hierarchy for a different mode ($M = 2$ for $m_+$ and $m_-$ in the BO case). In practice, we can set the termination elements $\hat{\rho}_{j_1 \cdots j_K}^{(n,m_-,m_+)}(t) = 0$ for $\gamma^{(n,m_-,m_+)} + \sum_{k=1}^{K} j_k v_k \gg \max\{\omega_0, \Delta\}$ and can reduce the number of hierarchy elements for calculations as $L_{calc} = L_{tot} - L_{term}$. Further inclusion of BO modes may be possible but the calculations become computationally expensive. In such case we my incorporate a variety of techniques developed for the HEOM approach to accelerate numerical calculations. [46,47,89,90,91,92,93]

## IV. LINEAR ABSORPTION AND TWO-DIMENSIONAL CORRELATION SPECTRA

In quantum mechanics, any physical observable is expressed as an expectation value of a physical operator. In an optical measurement, the observable at time $t$ is defined by $\langle \hat{\mu} \hat{\rho}_{tot}(t) \rangle$, where $\hat{\mu}$ is the dipole operator and $\hat{\rho}_{tot}(t)$ is the density matrix which depends on the interaction between the driving field and the system. [32] For linear absorption spectroscopy, the density matrix is expanded in terms of the laser interaction. If the laser interaction is expressed as $f(t)\hat{\mu}$, where $\hat{\mu} = |1\rangle\langle 0| + |0\rangle\langle 1|$, then the first-order expansion term for the impulsive excitation $f(t) = \delta(t - t_1)$ is expressed as

$$\begin{aligned} R^{(1)}(t_1) &= \frac{i}{\hbar} \langle [\hat{\mu}(t_1), \hat{\mu}(0)] \rangle \\ &= \frac{i}{\hbar} tr\left\{ \hat{\mu} e^{-i\hat{L}_{tot}^0 t_1} \hat{\mu}^{\times} \hat{\rho}_{tot}^{eq} \right\}. \end{aligned} \quad (4.1)$$

Here, $\hat{\rho}_{tot}^{eq} = e^{-\beta \hat{H}_{tot}^0} / tr\{e^{-\beta \hat{H}_{tot}^0}\}$ is the equilibrium state of the total system and we introduce the total Hamiltonian without the laser interaction $\hat{H}_{tot}^0$. The super-operators are defined by $e^{-i\hat{L}_{tot}^0 t_1} \hat{A} \equiv e^{-\frac{i}{\hbar}\hat{H}_{tot}^0 t} \hat{A} e^{\frac{i}{\hbar}\hat{H}_{tot}^0 t}$ and $\hat{A}^\times \hat{B} = \hat{A}\hat{B} - \hat{B}\hat{A}$, where $\hat{A}$, and $\hat{B}$ are ordinary operators. In 2D spectroscopy, the multibody correlation functions of a molecular dipole or polarizability are measured using ultra short pulses.[2][5] The third-order optical processes such as two-dimensional infrared and electronic spectroscopies are calculated for pulse sequences $f_1(t) = \delta(t - t_1)$, $f_2(t) = \delta(t - t_1 - t_2)$, and $f_3(t) = \delta(t - t_1 - t_2 - t_3)$ as[9][4]

$$R^{(3)}(t_3, t_2, t_1) = -\frac{i}{\hbar^3} \langle [[[\hat{\mu}(t_1 + t_2 + t_3), \hat{\mu}(t_1 + t_2)], \hat{\mu}(t_1)], \hat{\mu}(0)] \rangle$$
$$= -\frac{i}{\hbar^3} tr\{\hat{\mu} e^{-i\hat{L}_{tot}^0 t_3} \hat{\mu}^\times e^{-i\hat{L}_{tot}^0 t_2} \hat{\mu}^\times e^{-i\hat{L}_{tot}^0 t_1} \hat{\mu}^\times \hat{\rho}_{tot}^{eq}\}. \quad (4.2)$$

Here, $\hat{A}^\times \hat{B}^\times \hat{C} = \hat{A}(\hat{B}\hat{C} - \hat{C}\hat{B}) - (\hat{B}\hat{C} - \hat{C}\hat{B})\hat{A}$, where $\hat{A}$, $\hat{B}$ and $\hat{C}$ are ordinary operators. Since each $\hat{\mu}^\times$ can act either from the left or from the right, and since $R^{(3)}(t_3, t_2, t_1)$ contains three $\hat{\mu}^\times$, Eq.(4.2) naturally separates into eight terms. In practice we need to evaluate only half of these terms, since they always come in Hermitian conjugate pairs and we need four terms. Accordingly, the laser interactions are described by the transitions between the energy states, and the optical processes including the time ordering of the laser pulses are conveniently described by diagrams such as the double-sided Feynman (Liouville space) diagrams denote by (I)-(IV) in Fig 3.[2] By formulating third-order optical spectroscopy in Liouville space, it becomes possible to separate the process into three steps. The first pulse creates coherence during the $t_1$ period and then the second pulse brings the system in the population. There are actually two population states, one being the upper state $|1\rangle$ coming from paths (I) and (II), and the other being the lower state $|0\rangle$ coming from paths (III) and (IV). These two population states then propagate during the delay time $t_2$. In addition to the optical transitions obtained from coherent states, the population evolution is the significant information that is obtained from 2D spectroscopies. Finally, the system interacts with the third probe pulse and the signal for a single de-excited coherent state is generated in the $t_3$ period. Note that if the system consists of more than two levels, there is a contribution from double excitations corresponding to the $|0\rangle \rightarrow |1\rangle \rightarrow |2\rangle$ transition.[95,96,97]

By utilizing the phase matching condition, experimentalists can measure (I) and (IV), and (II) and (III) separately. Numerically such separation can be done by performing two-dimensional the Fourier transformation in time $t_1$ and $t_3$ as

$$I^{(3)}(\Omega_3, t_2, \Omega_1) = \int_0^\infty dt_1 \int_0^\infty dt_3 \, e^{-i\Omega_1 t_1} e^{i\Omega_3 t_3} R^{(3)}(t_3, t_2, t_1). \quad (4.3)$$

The first $(+\Omega_1, +\Omega_3)$ and the second $(-\Omega_1, +\Omega_3)$ quadrant of the Fourier plane correspond to pump-probe and photon echo spectra that arise from the diagrams (I) and (IV) and the diagrams (II) and (III), respectively.[94,95]

Using Eq. (3.6) we can evaluate the response function Eq.(4.2) as the following.[32,50,51,52] We first run the program sufficiently long period from a temporally initial condition (such as the factorized initial condition, $\hat{\rho}_{0\cdots 0}^{(0,0,0)}(0) = e^{-\beta \hat{H}_A^0}$ with the other hierarchy elements set to be zero) to have a true thermal equilibrium denoted by $\hat{\rho}_{j_1\cdots j_K}^{(n,m_-,m_+)}(t_\infty)$. Here and hereafter, $\hat{H}_A^0$ represents the system Hamiltonian without the laser interaction ($f(t)\hat{\sigma}_x = 0$). All of the hierarchy elements have to be utilized to define a correlated initial condition. Then the system is excited by the first dipole interaction $\hat{\mu}$ at $t=0$ as $\hat{\rho}'^{(n,m_-,m_+)}_{j_1\cdots j_K}(0) = \hat{\mu}^\times \hat{\rho}^{(n,m_-,m_+)}_{j_1\cdots j_K}(t_\infty)$. The perturbed elements $\hat{\rho}'^{(n,m_-,m_+)}_{j_1\cdots j_K}(t)$ then evolve in time by numerically integrating Eq.(3.6) with $\hat{H}_A^0$ up to $t=t_1$. At $t=t_1$, the system is excited by the second laser interaction as $\hat{\rho}''^{(n,m_-,m_+)}_{j_1\cdots j_K}(t_1) = \hat{\mu}^\times \hat{\rho}'^{(n,m_-,m_+)}_{j_1\cdots j_K}(t_1)$. After the distribution functions $\hat{\rho}''^{(n,m_-,m_+)}_{j_1\cdots j_K}(t)$ evolve in time with the initial condition $\hat{\rho}''^{(n,m_-,m_+)}_{j_1\cdots j_K}(t_1)$, at $t = t_1 + t_2$, the system is again excited by the third laser interaction as $\hat{\rho}'''^{(n,m_-,m_+)}_{j_1\cdots j_K}(t_1+t_2) = \hat{\mu}^\times \hat{\rho}''^{(n,m_-,m_+)}_{j_1\cdots j_K}(t_1+t_2)$. The elements $\hat{\rho}'''^{(n,m_-,m_+)}_{j_1\cdots j_K}(t)$ then evolve in time as $t = t_1 + t_2 + t_3$. Finally, the response function defined by Eq.(4.2) is calculated from the expectation value of the dipole moment as $R^{(3)}(t_3, t_2, t_1) = -itr\{\hat{\mu}\hat{\rho}'''^{(0,0,0)}_{0\cdots 0}(t_1+t_2+t_3)\}/\hbar$.[32] The linear absorption spectrum Eq.(4.1) can also be calculated in a same manner. Note that, to take into account the system-bath coherence (or system-bath entanglement [82,98]) during the external perturbation, it is important to operate $\hat{\mu}$ to all of the hierarchy elements $\hat{\rho}^{(n,m_-,m_+)}_{j_1\cdots j_K}(t)$. Although we only use $\hat{\rho}^{(0,0,0)}_{0\cdots 0}$ to calculate an expectation value, the other elements are essential to have an echo signal for a non-Markovian noise in 2D spectroscopy. [42,54,55]

It has been shown that population change through ET (or nonadiabatic) coupling can be explored by pump-probe spectroscopy [49,99] If we explore not only population dynamics but also the system-bath coherence (or system-bath entanglement) through the different pulse excitation, two-dimensional correlation spectra may be a better choice. This spectrum can be calculated from the first and second quadrant of the Fourier transformed response function $I_1^{(3)}(\Omega_3, t_2, \Omega_1)$ and $I_2^{(3)}(\Omega_3, t_2, \Omega_1)$ as

$$I_c(\Omega_3, t_2, \Omega_1) = \text{Re}\{I_1^{(3)}(\Omega_3, t_2, \Omega_1) + I_2^{(3)}(\Omega_3, t_2, -\Omega_1)\}, \tag{4.4}$$

and the signals are plotted as the function of $\Omega_1$ and $\Omega_3$ for different $t_2$. Note that since we consider the two-level system, the contribution from the double excitation which may be created by the third pulse for multilevel system does not exist and can be neglected. [95]

## V. NUMERICAL RESULTS

In the following, we set the excitation energy of the system as the base unit $\omega_0 = 1$ and calculated linear absorption spectrum and 2D correlation spectrum for different ET coupling strength $\Delta = 0.0\,\omega_0$, $0.2\,\omega_0$, and $0.4\,\omega_0$. The bath parameters were $\beta\hbar = 2.0/\omega_0$, $\lambda_o = 0.01\omega_0$, $\gamma_o = 0.1\omega_0$, $\gamma_u = 0.01\omega_0$, $\omega_u = 0.2\omega_0$ and we consider the two cases of the displacement (reorganization energies) (A) $\lambda_u = 0.05\omega_0$ and (B) $\lambda_u = 0.2\omega_0$. The temperature here we considered is very high for a case between the lower state and upper state transition. To lower the temperature ($\beta\hbar\omega_0 \approx 10$), one may need to employ numerical techniques to accelerate calculations.[89-93] It should be noted that dynamical behavior of the system does not change so much once the temperature becomes low enough compared with the characteristic frequency of the system (such as $\omega_u$), since the thermal excitation becomes so small that the fluctuation does not play an any role for electronic excitation at very low temperature. So, in practice, we do not have to lower the temperature below $\beta\hbar\omega_u \approx 10$ for $\omega_0 \gg \omega_u$.

The case for $\Delta = 0.0$ can be calculated analytically using the response function approach[26-28] even higher than the third-order response,[29,30] the other cases are almost impossible to study from other approaches including the TCL Readfield approaches, since, to have right 2D profiles, we have to deal with a non-Markovian noise with a strong-system bath coupling characterized by $\lambda_o/\gamma_o$ and $\lambda_u/\gamma_u$ with including a system-bath coherence that cannot be neglect to calculate multidimensional spectroscopies.[42]

The numerical integrations of the hierarchy equations of motion were performed by using the 4th-order Runge-Kutta method. We chose $N = 12 \sim 14$ and $K = 4$, so the total numbers of the hierarchy elements used for calculations are $L_{calc} = 31823 \sim 38759$. The profiles of calculated distribution function $J'(\omega)$ for the case (A) and (B) are depicted in Fig. 4.

The Fourier transformed linear absorption spectra calculated from Eq.(4.1) are presented in Fig. 5. Each of the side peaks represents transitions between the vibrational levels of the lower and upper electronic states. If we denote the lower and upper states vibrational levels by $n$ and $m'$, where $n$ and $m'$ are the integer, the peak at around $\Omega_1/\omega_0 = 0.6, 0.8, 1.0, 1.2$, and $1.4$ in Fig. 5(A), for example, corresponds to the $2 \to 0'$, $1 \to 0'$, $0 \to 0'$, $0 \to 1'$ and $0 \to 2'$ transitions, respectively. Note that the peak at $\Omega_1/\omega_0 = 1.0$ also involves a contribution from $1 \to 1'$ and $2 \to 2'$ transitions, since the temperature is higher than the vibrational excitation energy. Since the system is initially in the thermal equilibrium state, the populations in the lower vibrational levels are higher. Thus the $1 \to 0'$ peak is lower than the $0 \to 1'$ peak. The contribution from the Drude spectral distribution is observed as a broadened Gaussian peak under the vibrational transition peaks. As can be seen from Eq.(2.8), the displacement of oscillators between the $|0\rangle$ and $|1\rangle$ states is determined by $\lambda_u$. In Fig.5 (B) for large $\lambda_u$, we observe many phonon peaks due to the varieties of the vibrational transitions arises from the large displacement of the potential surfaces.

When the ET coupling $\Delta$ becomes stronger, the spectrum shifts to the blue because the energy levels are defined by $\pm\sqrt{\omega_0^2 + \Delta^2}/2$. Besides this blue shift, the effects of the ET coupling are negligible. This is because the linear absorption process does not involve the population state $|1\rangle\langle 1|$ as shown in the $t_2$ period of the third-order response in Fig. 3. The linear absorption measurement can only detect the coherence between the $|0\rangle$ and $|1\rangle$ states.

The calculated 2D correlation spectra in the case (A) and (B) for different $t_2$ are presented in Figs. 6 and 7, respectively. In each of the figures, the diagonal peaks around $(\Omega_1, \Omega_3) = (0.8, 0.8), (1.0, 1.0), (1.2, 1.2)$, and so on in Fig. 6 and $(\Omega_1, \Omega_3) = (0.7, 0.7), (1.0, 1.0), (1.3, 1.3)$ and so on in Fig. 7 corresponds to the $1 \to 0' \to 1$, $0 \to 0' \to 0$, $1 \to 0' \to 1$ transitions and so on, while the off-diagonal peaks such as $(\Omega_1, \Omega_3) = (1.0, 0.8), (1.0, 1.2)$ in Fig. 6 correspond to $0 \to 0' \to 1$, $1 \to 1' \to 0$, and so on.

At time $t_2 = 0$ in Figs. 6 (i-a)-(i-c) and Figs. 7(i-a)-(i-c), only the diagonal peaks are prominent, since there is not enough time for the excited wave packets to decay. At time $t_2 = 5$ in Figs. 6(ii-a)-(ii-c) and Fig. 6(ii-a)-(ii-c), the peak profiles become a symmetrical cross like shape. This is because the coherence between the ground and excited states vibrational motions are lost quickly in the present parameter regime and, as a result, the present 2D signal exhibits the uncorrelated transitions in the $t_1$ period such as $0 \to 0'$, $1 \to 0', 0 \to 1'$ and in the $t_3$ period such as $0' \to 1$, $1' \to 0$ etc.[95,96] In Figs. 6(iii) and (iv) and 7 (iii) and (iv), the height of each phone peak slightly change in time due to the movement of the wavepacket created in the upper potential surface. The movement of wavepacket itself can be observed explicitly if we include the coordinates $q_j$ in the calculations.[49,99]

In the present model, the ET transition between the electronic states as well as relaxation between the phonon bands takes place. The relaxation of phonon bands is much faster than the ET process, thus, at the early stage of the $t_2$ evolution, the profiles of phone peaks are similar regardless of the ET coupling. In the case (A) $\lambda_u = 0.05$, the energy barrier between the two potential surfaces is high due to the small displacement and the ET transition only occurs at higher vibrational levels in the excited state potential, while, in the case (B) $\lambda_u = 0.2$, the energy barrier is low and the ET transition occurs at lower vibrational levels. Thus, while the signals exhibit the similarity in Figs. 6 (ii)-(iv), the lower frequency peaks in the $\Omega_3$ direction are suppressed in Fig. 7 (ii)-(iv) if the ET coupling becomes strong, since the lower vibrational excitations in the $|1\rangle$ state vanishes due to the ET transition. In Fig. 7, the entire peak volume decreases as $t_2$ increases for large ET coupling, while, in Fig. 6, the decrease of the volume is small due to the large energy barrier.

A clear indication of ET coupling in Figs. 6 and 7 is observed as the peaks at $\Omega_3 = 0$ spreads on the $\Omega_1$ axis. To illustrate the outline of these narrow peaks, we replot Fig. 7 (ii-c) as the 3D picture in Fig. 8 as an example. The narrow peaks along $\Omega_3 = 0$ at phonon-band positions are observed. The existence of these peaks can be easily understood if we regard the ET coupling $\Delta$ as a laser interaction with frequency 0. These peaks do not appear on the $\Omega_3$ axis, since $\omega_0$ is very large compared with the thermal activation energy and there is no $|0\rangle \rightarrow |1\rangle$ transition without the pump excitation. The existence of the $\Omega_3 = 0$ peaks indicate that if the ET coupling is time-dependent due to some other degrees of freedom like in the case of proton-coupled electron transfer,[100,101,102] then we may monitor that time-dependence from the peak profile. Since ET coupling is in a same form as a laser interaction, we may also investigate the ET transition induced by Stark effects by strong laser in a same manor.[103,104,105]

VI. CONCLUSIONS

We have analyzed the ET process of a two-level system coupled to an overdamped Drude and underdamped Brownian oscillators using equations of motion that allowed us to incorporate two dephasing modes at finite temperature. Although 2D correlation spectrum is also based on the third-order response function like the pump-probe, hole burning, and photon echo, we demonstrated that we can subtract the information for the ET coupling and relaxation process by analyzing a 2D signal as the peaks along the $\Omega_3 = 0$ and the decrease of the total peak volume in time $t_2$. For a large displacement case, we also observe the suppression of the lower phonon sideband peaks due to the ET transition.

Here, we analyze 2D spectrum for a two-level system in a limited parameter regime, but an extension to a multi-level system for a realistic parameter set corresponds to the ET transition in a reaction center is possible.

[63,64] The present approach can also be applied to a system driven by pulses of arbitrary number, shape, and strength, as well as a system with time-dependent ET couplings.[62] The present formulation can also be extended to multimode Brownian oscillator systems by introducing a higher dimensional hierarchy. Inclusions of multi-mode are computationally very expensive, and therefore one has to employ a variety of numerical techniques developed for HEOM approach to accelerate numerical calculations.[46,47,89,90,91,92,93] Here, we assumed that the primary oscillator modes are harmonic. However, if we employ a less reduced density matrix in which the bath ($x_{n_j}$) modes are eliminated and we still keep the oscillator coordinates $q_j$, we can relax this limitation. The density matrix can then be described as a wavepacket in phase space by using the Wigner representation.[44,48,49,99]

## ACKNOWLEDGMENTS

The author is grateful for the hospitality of Professor Dwayne J. Miller and his group members in DESY, Hamburg. The financial support from Humboldt Foundation and a Grant-in-Aid for Scientific Research B2235006 from the Japan Society for the Promotion of Science are acknowledged.

FIG. 1: Potential surfaces of the linearly displaced harmonic oscillator system. The lower state is denoted $|0\rangle$, whereas the upper is $|1\rangle$. The equilibrium coordinate displacement, the ET coupling, the oscillator frequency, and the energy difference between two potentials are expressed by $d_j$, $\Delta_j$, $\omega_j$, and $\omega_0$, respectively. Red and blue represents the pump excitation and probe de-excitation with frequencies $\Omega_1$ and $\Omega_3$, respectively.

FIG. 2: Schematic view of the system-oscillator-bath coupling. In the case (a), the two oscillators are independently coupled to their own bath, whereas, in the case (b), one oscillator coupled to two baths. In the present Drude+BO model, the cases (a) and (b) become identical.

FIG.3: Double-sided Feynman diagrams of the third-order response function. The left and right lines represent the time evolution of the left (ket) and the right (bra) hand side of the density matrix, respectively. The thin blue and the thick red lines denote the lower state $|0\rangle$ or $\langle 0|$ and the upper state $|1\rangle$ or $\langle 1|$. The paths (I)-(IV) correspond to the process $|0\rangle\langle 0|$ to $|0\rangle\langle 0|$. The Hermitian conjugate paths which can be obtained by interchanging the left and right lines, respectively, are not shown here.

FIG. 4: Spectral distribution $J'(\omega)$ defined by Eq.(3.3) is plotted in (A) the small displacement case $\lambda_u = 0.05\omega_0$ (blue) and (B) the large displacement case $\lambda_u = 0.2\omega_0$ (red) for the parameters $\lambda_o = 0.01\omega_0$, $\gamma_o = 0.1\omega_0$, $\gamma_u = 0.01\omega_0$, and $\omega_u = 0.2\omega_0$.

FIG. 5: Absorption spectrum (Eq. (4.1)) plotted in (A) the small displacement case $\lambda_u = 0.05\omega_0$ and (B) the large displacement case $\lambda_u = 0.2\omega_0$ for different ET couplings $\Delta = 0.0\omega_0$ (green), $0.2\omega_0$ (red) and $0.4\omega_0$ (blue), respectively.

FIG 6: Two-dimensional correlation spectrum $I_c(\Omega_3, t_2, \Omega_1)$ for different values of $t_2$ and different ET couplings $\Delta$ in (A) the small displacement case $\lambda_u = 0.05\omega_0$. We plot (a) $\Delta = 0.0\omega_0$, (b) $\Delta = 0.2\omega_0$, and (c) $\Delta = 0.4\omega_0$ at different times (i) $t_2 = 0$, (ii) $t_2 = 5$, (iii) $t_2 = 10$, and (iv) $t_2 = 20$, respectively. The scale of the signal intensity is chosen to be the same. The peaks at $\Omega_3 = 0$ spreads on the $\Omega_1$ axis arise from the ET coupling in the case (b) and (c).

FIG 7: Two-dimensional correlation spectrum $I_c(\Omega_3, t_2, \Omega_1)$ for different values of $t_2$ and different ET couplings $\Delta$ in (B) the large displacement case $\lambda_u = 0.2\omega_0$. The other parameters are the same as the case in Fig. 6. The peaks at $\Omega_3 = 0$ spreads on the $\Omega_1$ axis arise from the ET coupling in the case (b) and (c).

FIG 8: Three dimensional profile of the two-dimensional correlation spectrum presented in Fig. 7 (ii-c). The narrow peaks along $\Omega_3 = 0$ at phonon-band positions arise from the ET transition.

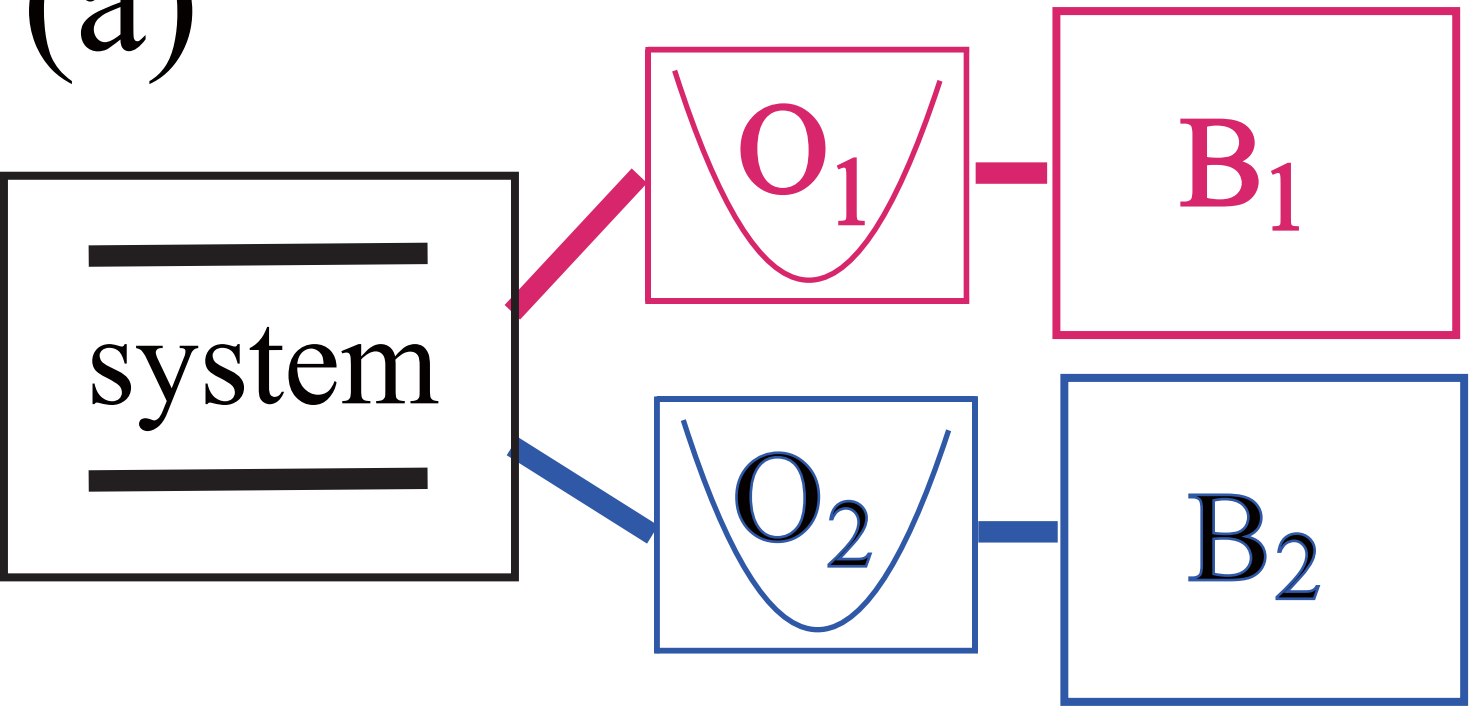
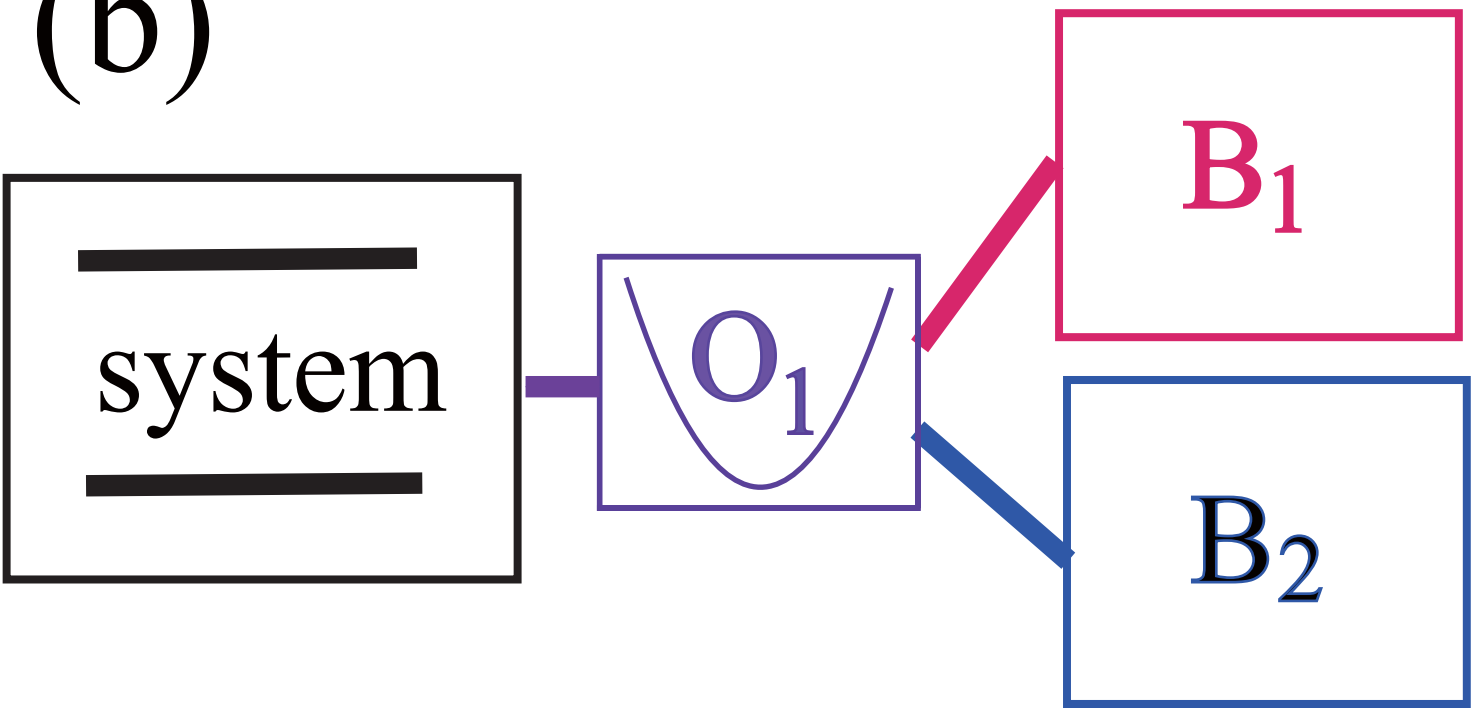

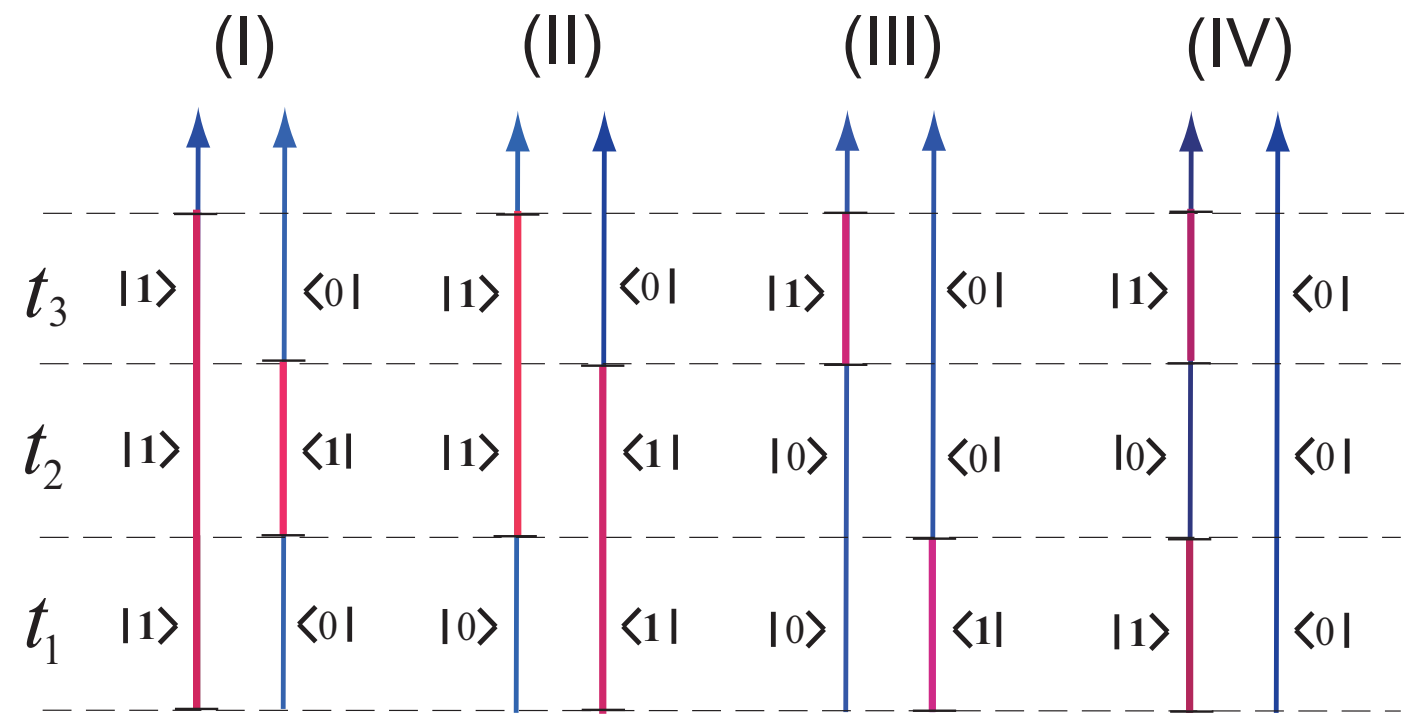

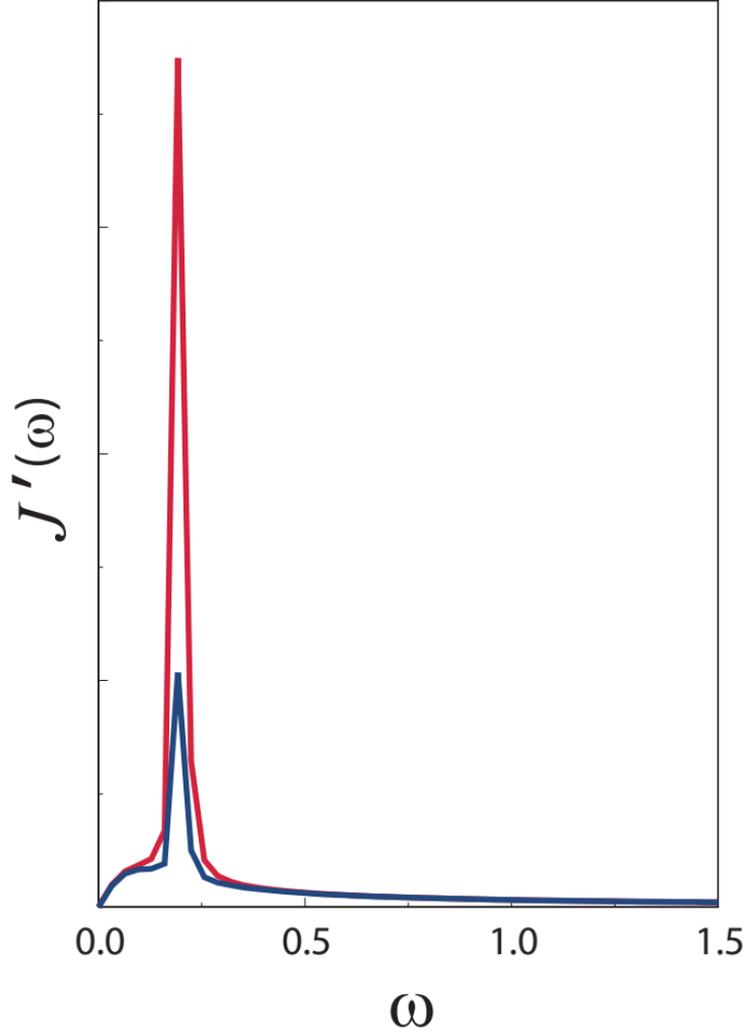

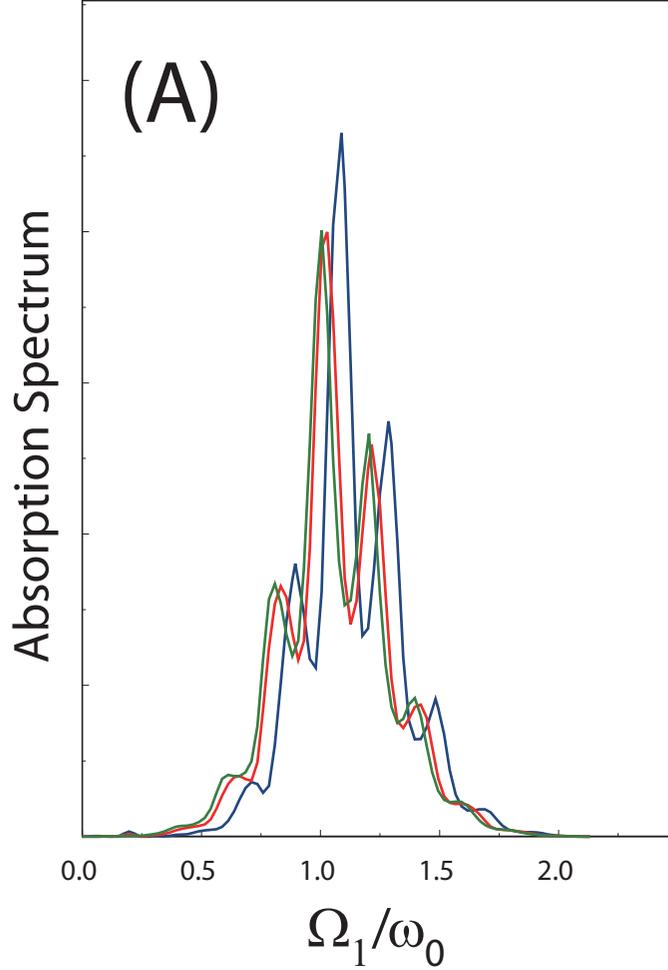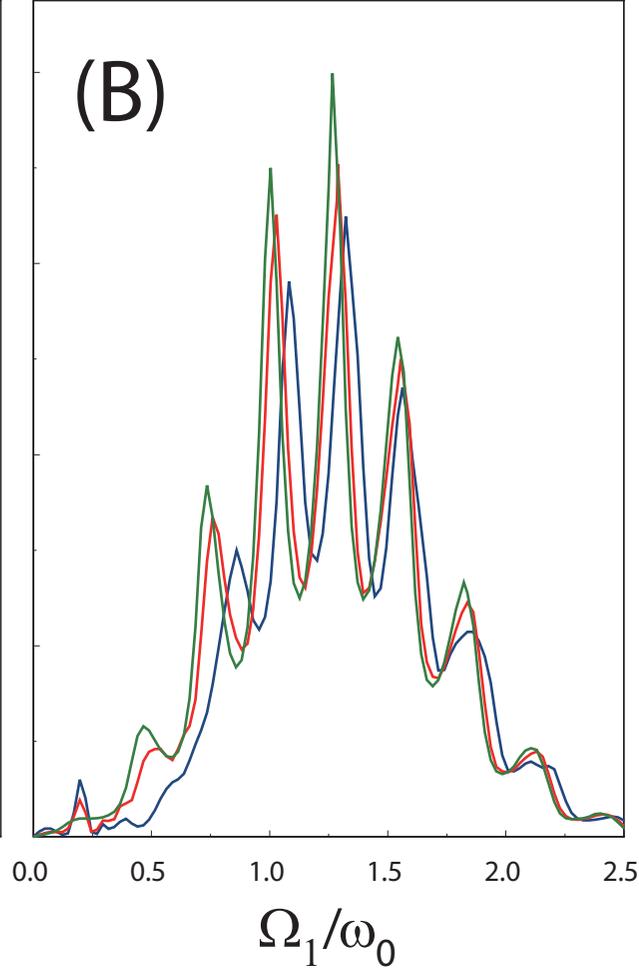

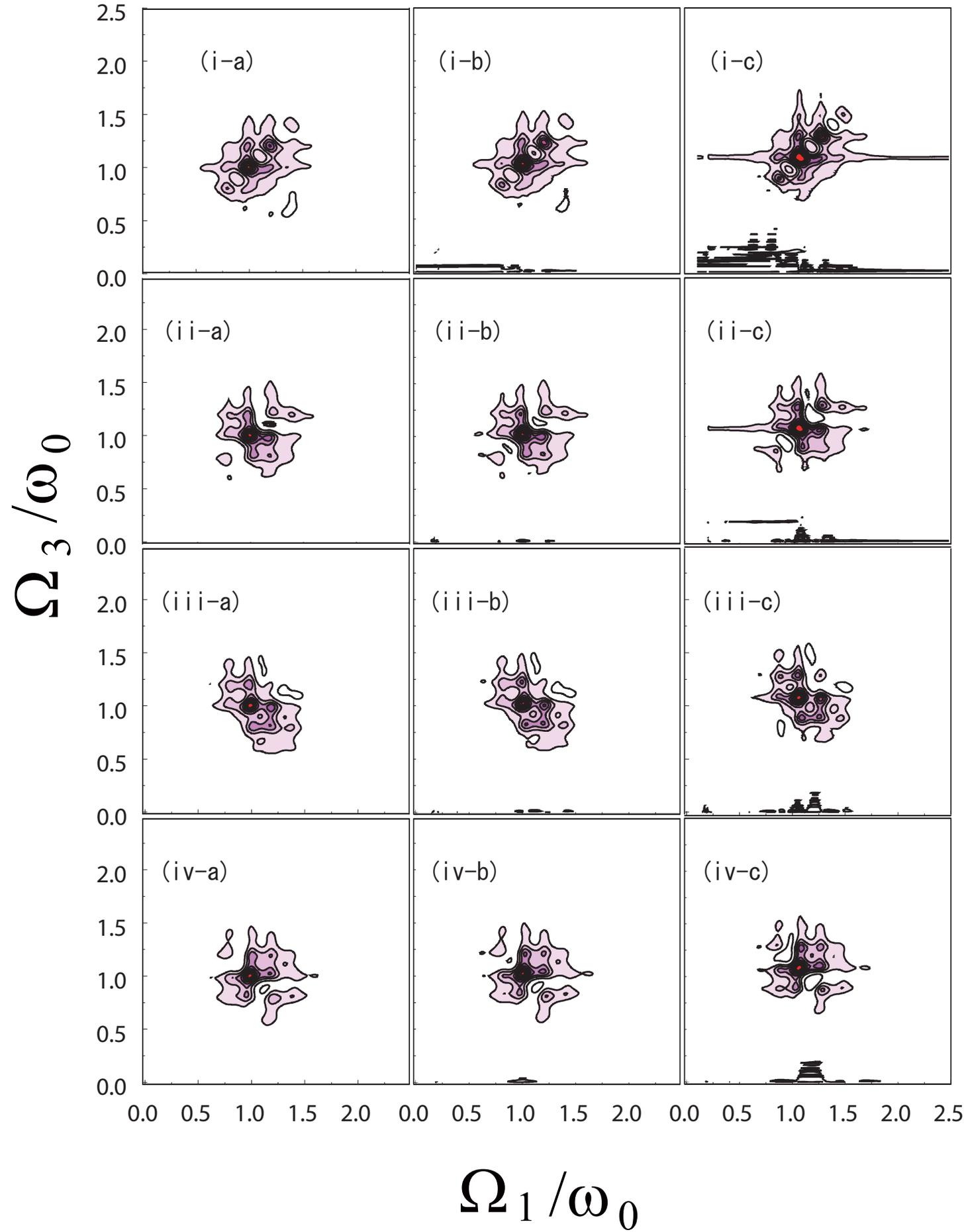

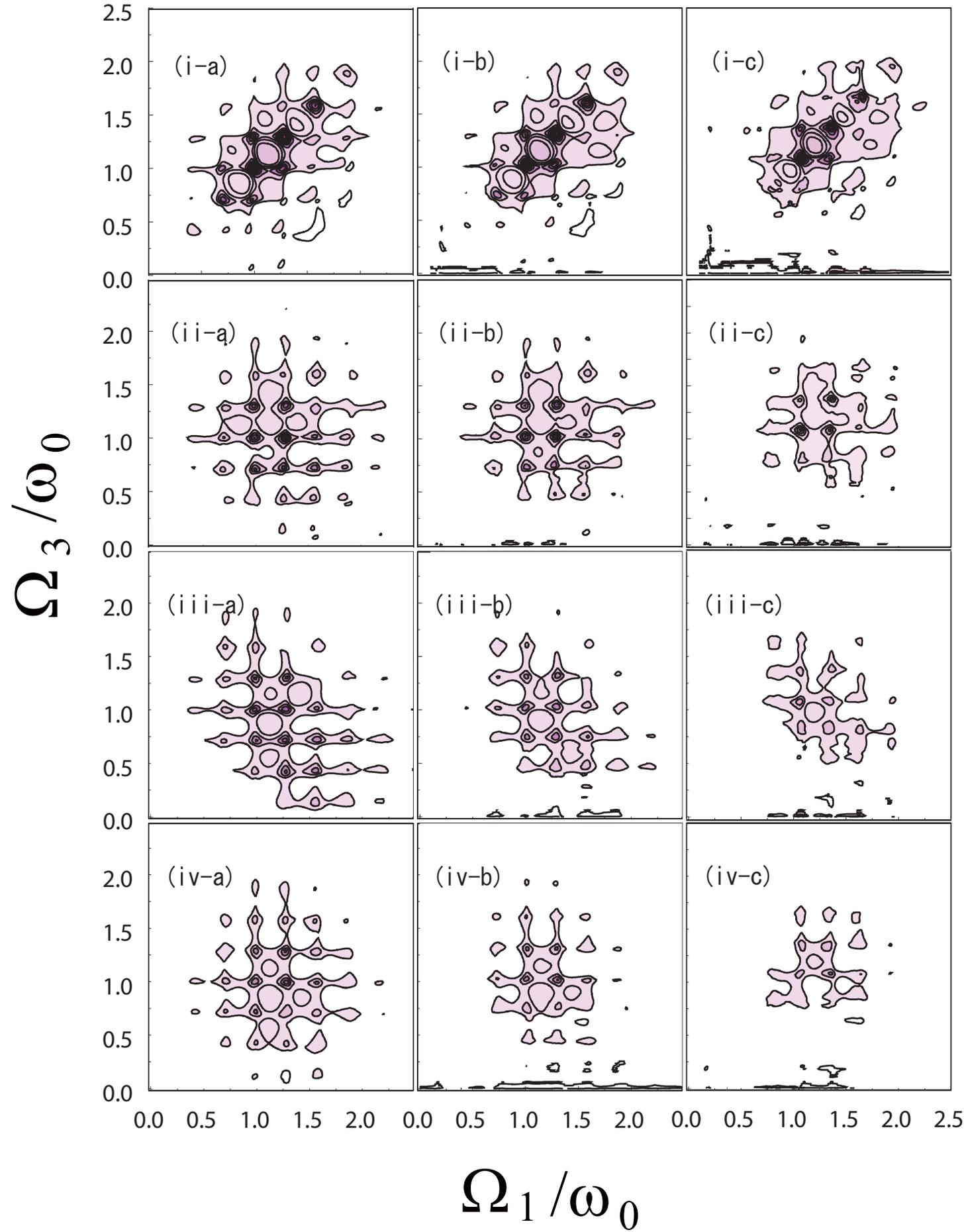

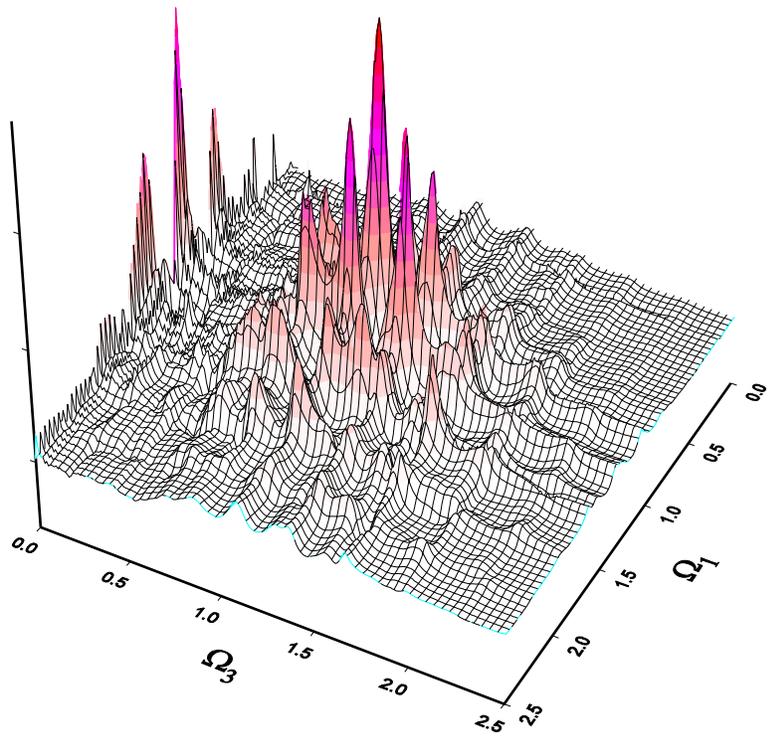